\documentstyle[prb,aps,preprint]{revtex}
\newcommand{\no}{\nonumber}
\newcommand{\be}{\begin{equation}}
\newcommand{\ee}{\end{equation}}
\newcommand{\beq}{\begin{eqnarray}}
\newcommand{\eeq}{\end{eqnarray}}
\begin{document}
\draft
\title{Crossover from Spin-Density-Wave to Neel-like Ground State}
\author{Jaime Ferrer}
\address{Departamento de F\'{\i}sica de la Materia Condensada,
Universidad Aut\'onoma de Madrid, 28049 Madrid, Spain}
\maketitle
\begin{abstract}
The characterization and evolution of a Spin Density Wave into the
Quantum Neel ground state is considered in the context of a weak
coupling theory of the half-filled Hubbard model. Magnetic
properties obtained from this weak coupling approach in one
dimension compare
favorably with exact results from Bethe ansatz (BA). A study of
the evolution of several length scales from weak to strong
coupling is also presented.
\end{abstract}


\section{Introduction}
Interest in low-dimensional Quantum Antiferromagnetism experienced
a sharp rise with the
realization that magnetic fluctuations are at the root of
many of the
exotic properties of High Temperature Superconductors.
It is fairly well established by now that the plain two-dimensional
Quantum Heisenberg
model can account for most of the experimental features of undoped
cuprates
\cite{keimer,chn,subir1}. It is also widely believed that the simple
one-band Hubbard Model (HM) can map large parts of the
experimental phase diagram
\cite{keimer,anderson,pines,andrey1}, and most of the theoretical
work has used it as an appropriate starting point
\cite{chn,subir1}.
Simple as it looks, the HM displays a rich variety of regimes, with
highly non-trivial physics. In fact, there is no consensus about
the adequate low-energy effective action for
small doping concentrations
\cite{siggia,millis,subir2}, nor about
the gross features of the phase
diagram in that regime \cite{siggia,schulz1,shastry}.

The situation is much more
clear exactly at half-filling, because charge
degrees of freedom are frozen out.
In this case and for large Coulomb repulsion, the physical
electrons are localized,
and the ground state is the quantum analog of the Neel
state. In the
opposite limit of small $U$, the ground state is a
Spin Density Wave (SDW), and should be adequately described using
standard RPA
theory over a broken symmetry, Mean Field SDW state.
In an early development, Schrieffer, Wen and Zhang
\cite{swz} proposed
that the later, weak coupling, approach (to be called RPA-SDW
hereafter) could also explain the physics
of the Quantum Neel state. Later on, Chubukov and Frenkel
\cite{andrey2} explicitly showed that many of the known
results for the
Heisenberg model could
indeed be recovered within the SDW-RPA approach.
There are, nevertheless,
some details missing in the physical picture of the
half-filled HM. For instance: it is common lore that a SDW is the
analog of the Neel state when the electrons are itinerant, but a
clear
and quantitative description of it is lacking. As a direct
consequence, the crossover regime between both states is not
fully understood. The author also feels that there is some
confusion about the Temperature and length scales generated
in the different
regimes. To be specific: these scales are, in general, well
established
for the positive-$U$ HM at strong coupling and for the negative-$U$
case
at weak coupling, but not in the other two limits \cite{randeria}.

The purpose of this paper is to shed some more light on these
issues.
In order to do so, I will apply the RPA-SDW
expansion used in ref. 12 to the one-dimensional
case at Zero Temperature and compare with the Bethe ansatz solution
\cite{lieb}. Hence, the whole phase diagram at half-filling will be
covered. I will also  draw many consequences on the scales involved
in
the problem by the heavy use of duality relations \cite{shiba1}.
The computations are performed for the worst
case,
because RPA schemes are supposed to do a bad job in
low
dimensions and in the spin-disordered phase. But I will
actually show that RPA
works remarkably well even in 1d -at half filling-.

Some comments are in order at this point of the introduction.
The HM at half-filling falls into the O(3), instead of XY,
universality class.
A closely related fact is that its ground state
is a singlet in the strong coupling limit \cite{mattis} so that
all correlation functions are rotationally invariant.
A further, specific feature of one-dimensional
systems
is that they do not possess Long Range Order, even at zero
Temperature, because of thermal and
quantum fluctuations \cite{mermin}. Correlation functions related
with the order parameter have a mass gap and do not follow
Goldstone behavior \cite{haldane1}. The complications derived from
spin-charge separation are eluded because charge degrees of freedom
are frozen out at half-filling \cite{lieb}.

The layout of this paper is as follows: several magnitudes will be
computed within RPA-SDW and compared with Bethe ansatz results in
section
2. Section 3 will present a
study of the different length and Temperature
scales of the problem. A brief conclusion will end the paper.
Details
of the calculations will be relegated to appendices A and B.
Energies will be measured in units of the tunneling amplitude, $t$.

\section{Study of the SDW-RPA ansatz at $T=0$}
It is commonly believed that RPA approaches should not perform
well in
low dimensions or when one is trying to describe physical properties
of the strong coupling phase. The purpose of this section
is to show that RPA-SDW
produces accurate
results at either weak or strong coupling even in the
extreme case of one dimension, as long as the ground state is
close enough to a commensurate Antiferromagnet.

The first quantity which arises in RPA-SDW theory is the magnitude
of the Mean Field charge gap, $2\,\Delta$, which is obtained from
the equation
\be
\frac{1}{U}=\frac{1}{N}\sum_k \! '\,\frac{1}{E_k}\no
\ee
Inclusion of RPA fluctuations does not modify the equation.
Although the asymptotic
expression for $\Delta$ when $U$ approaches zero, $\frac{8}{\pi} U
e^{-2\pi/U}$, is not captured by the RPA-SDW approximation,
which gives $\Delta\sim e^{-2\pi/\sqrt{u}}$,
the numerical differences
are negligible for such small values of $U$ (see fig. 1).
The RPA-SDW curve begins to depart appreciably from the exact
result for values of $U$ of the order of the bandwidth, $D$.
This fact suggests that the crossover
from nearly itinerant to localized
electron behavior is not described quantitatively within this
weak-coupling approximation.

The exact large-$U$ value of the charge gap, on the other hand,
is $U-4 t = U - D$,
where $D$ is the free band width \cite{ovchinnikov}.
RPA-SDW theory overestimates $2\,\Delta$ by $D$, because
the approximate
lower and upper Hubbard bands tend to $\delta$-fuctions
while the exact ones retain a bandwidth of $D$
when $U\rightarrow\infty$ (see fig. 1 again).

The collective excitations of the system appear as poles of the
imaginary part of the
different correlation functions $\chi"(q,w)$. The weight of a
particular fluctuation $(q,w)$ on the ground state is given by
the residue of $\chi"$ at that particular value of $(q,w)$.
The RPA-SDW ansatz, in
conjunction with Goldstone's theorem, tells us that the only
low-energy collective excitations of the half-filled HM at zero
Temperature and for large $U$ are spin waves, and show up as
poles of the transverse spin susceptibility,
$\chi^{+-}_{RPA}"$. Charge and longitudinal spin
fluctuations correspond to poles of $\chi^{00}_{RPA}"$
and $\chi^{zz}_{RPA}"$, and have a mass gap. These correlation
functions have a very simple form in the strong coupling limit.
For instance
\be
\chi_{RPA}^{+\,-}(q,w) = \frac{1}{U}\:\frac{w^2-w_q^2+4
(1-\cos(q))}{2\,w_q}
\left\{\frac{1}{w+w_q+i\delta}-\frac{1}{w-w_q+i\delta} \right\}
\ee

Fig. 2 (a)
shows the imaginary part of the transverse spin susceptibility
for $U=20$.
It  displays the typical Goldstone-mode structure and
has a strong q-dependence, proper for  spin waves.
Notice also that the weight of the Ferromagnetic zero mode
tends to zero, while that of the Antiferromagnetic one diverges,
in agreement with strong coupling analyses \cite{haldane2}.
On the contrary, $\chi^{zz}_{RPA}"(q,w)$ for $U=20$ is almost
dispersionless and has a large gap of order $U$ (see figure 2 (b)).
Hence, longitudinal spin fluctuations are massive and possess a
localized nature.

The exact spin correlation functions
should be rotationally invariant
because the ground state of the system is a singlet at strong
coupling. Regretfully, the RPA-SDW
response functions do not have this property.

Returning to the strong-coupling case again, it is easy to show
that the excitation spectrum given by RPA-SDW is
\be
w_q = v_s |\sin(q)| \sim_{q\rightarrow 0} v_s q
\ee
This is the result one would obtain for the Heisenberg model
in Linear Spin Wave theory (LSWT) \cite{mattis}.
The linear relation between $w$ and
$q$ for small $q$ holds all the way down to $U=0$ and serves to
define $v_s$ for all values  of the Coulomb repulsion (fig. 3).
Notice that the RPA-SDW velocity grows as
$U$ increases from zero and then has a
maximum for $U\sim 1$. What happens here is that the limit
$U\rightarrow 0$, where $v_s=2$, does not coincide with the $U=0$
case ($v_s=0$). This subtle point is well captured by the exact
BA analysis, but not by RPA-SDW theory, where $v_s$ begins at
zero, increases steeply until it reaches a value close to
$3$ at the maximum and then approaches the BA curve for
for $U\sim D$. Both curves
decrease approximately at the same pace for even larger $U$ and
tend asymptotically
to $\frac{\pi}{2} J\sim 1.6 J$ and $J$. The value given by RPA-SDW
theory for $v_s$ in higher dimensions ($\sqrt2 J$ for d=2) coincides
with that obtained from LSWT.

The squared magnetic moment, $<S^2>$ is displayed in
fig. 4, as given by Mean Field Theory (dashed line) and
Bethe ansatz (solid line). It
is quite remarkable the fact that Mean Field
Theory already  give the right
$U\rightarrow 0$ and $U\rightarrow \infty$ limits. Notice also
how Mean Field Theory underestimates grossly $<S^2>$
for small values of $U$, while
tends correctly to $3/4$ in the opposite limit.
What this seems to imply again is
that the Mean Field spin degrees of freedom are
{\em too itinerant} in the crossover regime from itinerant to
localized electron behavior.

RPA fluctuations do not cure this flaw. In fact,
$<S^2>$ diverges at strong coupling as
\be
<S^2>=\frac{1}{4} \sum_{k}\,'\:\frac{1-\cos(k)}{\sin(k)}
\rightarrow \infty
\ee
which is also the result obtained by LSWT.
In the opposite limit, on the contrary,
RPA-SDW theory reproduces the exact free result $\frac{3}{8}$.

Another important result concerns the average of the staggered
magnetization for large $U$, which also coincides
with LSWT   \cite{andrey2}
\be
<S^z>=\frac{1}{2}\left\{1-2 \sum_{k}\,'\:\frac{1-\sin(k)}{|\sin(k)|}
\right\}\rightarrow -\infty
\ee
This result implies that one can obtain a charge insulating
spin disordered ground state, also called {\em spin liquid},
 within SDW-RPA theory, because the
charge gap is still given by the Mean Field result.
LSWT is known to diverge in one dimension due to infrared
divergences, but to lead to amazingly accurate results in
higher dimensions \cite{igarashi,ferrer}.

Figure 5 is a useful eye-guide to understand the differences
between the Neel state and a Spin Density Wave. For large values
of $U$, the charge degrees of freedom of physical electrons are
frozen out and the spin degrees of freedom become localized
magnetic moments: the Heisenberg Hamiltonian provides a good
description of the HM in this limit. In this case, $<S^2> = S
(S+1)=3/4$. In the opposite case, $<S^2>$ equals $3/8$, which is
the value for free electrons. As soon as $U$ differs from zero, a
charge gap opens up, and the expectation value of the staggered
magnetization --in dimensions higher than one-- becomes finite
albeit small. The squared magnetic moment increases with $U$ from
$3/8$ to $3/4$ continuously but with a steep slope. By the time
when $U$ equals the band width, $<S^2>$ is already very close to
$3/4$, and the Spin Density Wave has fully developed  into the
Neel state. Likewise, for that value of $U$, the exact spin wave
velocity almost coincides with $\frac{\pi}{2} J$.

\section{Length scales in the Hubbard Model}
The transformation
\begin{eqnarray}
\hat{f}_{i\uparrow}^+&
\Leftrightarrow &\hat{f}_{i\uparrow}^+\nonumber\\
\hat{f}_{i\downarrow}^+&\Leftrightarrow &e^{i \bf{Q}{\bf R_{i}}}
\hat{f}_{i\uparrow}^+
\end{eqnarray}
where ${\bf Q}=(\pi,\pi,\dots)$, changes the positive-U half
filled HM into the negative one, and vice versa. It also
exchanges spin and isospin operators \cite{shiba1}:
\begin{eqnarray}
\hat{\cal{H}} = - \,t \sum_{<i,j>,\sigma} \hat{f}_{i,\sigma}^+
\hat{f}_{j,\sigma}+
U \sum_i \hat{n}_{i\uparrow}\hat{n}_{i\downarrow}
&\Leftrightarrow&
\hat{\cal{H}} = - \,t \sum_{<i,j>,\sigma} \hat{f}_{i,\sigma}^+
\hat{f}_{j,\sigma}-
U \sum_i \hat{n}_{i\uparrow}\hat{n}_{i\downarrow}\no\\\no\\\no\\
\left\{\frac{1}{2} (\hat{n}_{i\uparrow}-\hat{n}_{i\downarrow}),
e^{-i{\bf Q R_i}}
\hat{f}_{i\uparrow}^+\hat{f}_{i\downarrow},
e^{i{\bf Q R_i}}
\hat{f}_{i\downarrow}^+\hat{f}_{i\uparrow}\right\}
&\Leftrightarrow&\left\{\frac{1}{2}(\hat{n}_i-1),
\hat{f}_{i\uparrow}^+\hat{f}_{i\downarrow}^+,
\hat{f}_{i\downarrow}\hat{f}_{i\uparrow}\right\}\no\\\no\\\no\\
(\hat{S}^z, e^{-i{\bf Q R_i}} \hat{S}^+,
e^{i{\bf Q R_i}} \hat{S}^-)&\Leftrightarrow&
(\hat{\psi}^z,\hat{\psi}^+,\hat{\psi})
\end{eqnarray}

These duality relations allow  to infer many properties of one
model from the other. For instance,
the positive-$U$ HM at half-filling possesses a global
$SU(2)_{sp}\otimes U(1)_{ch}$. This means that its counterpart
has a global
$SU(2)_{ch}\otimes U(1)_{sp}$.
Likewise, an
\begin{itemize}
\item  Antiferromagnet (AFM)
polarized along the Z-axis is equivalent to a
commensurate Charge Density Wave
\item  AFM polarized along the XY-plane is equivalent to a
uniform superconductor
\item  Ferromagnet (FM) polarized along the Z-axis is equivalent
to the uniform state
\item  FM polarized along the XY-plane is equivalent to a
commensurate {\em staggered} superconductor \cite{fulde}
\end{itemize}

Because the ground state of an AFM at strong coupling is a
singlet, a uniform superconductor should be degenerate with a
CDW for large $U$.
Conversely,  the  spin transport properties of an AFM
polarized along the Z-axis and XY-plane should be very different,
because the charge transport properties of a uniform
superconductor and a commensurate CDW indeed are.
The appearance of the nesting wavevector ${\bf Q}$ in the duality
relations introduces, however, a slight difference because
staggering means folding of the Brillouin
zone into two pieces, with the corresponding doubling of the
periodicity.
It is also interesting to notice  that there do not seem to
be many experimental realizations of an staggered superconductor,
while a uniform FM polarized in the XY-plane takes place in
Nature very frequently \cite{caveat}.

The remaining of this  section is devoted to elucidate
what are the different
length scales which appear
in the positive and negative-U HM and how
they show up in the following correlation functions:
\begin{eqnarray}
\tilde{G}_{SDW}(k,k+Q) &=& \left(\begin{array}{cc}
<T \hat{f}_\sigma(k) \hat{f}^+_\sigma(k)>&
<T \hat{f}_\downarrow(k) \hat{f}^+_\uparrow(k+Q)> \\
<T \hat{f}_\uparrow(k+Q) \hat{f}^+_\downarrow(k)>&
<T \hat{f}_\sigma(k+Q) \hat{f}^+_\sigma(k+Q)>
\end{array}\right)
\Leftrightarrow \nonumber\\&&\nonumber\\&&
\tilde{G}_{sup}(k)
=\left(\begin{array}{cc}G_0 &
F\\F^+&G_0\end{array} \right)
=\left(\begin{array}{cc}
<T \hat{f}_\sigma(k) \hat{f}^+_\sigma(k)>&
<T \hat{\psi}(k)> \\
<T \hat{\psi}^+(k)>&
<T \hat{f}^+_{-\sigma}(k) \hat{f}^+_{-\sigma}(k)>
\end{array}\right)\nonumber
\end{eqnarray}
\be
C^{+-}(k)\,=\,<T\,S^+(k)\,S^-(-k)>\Leftrightarrow
C_{\psi}(k)\,=\,<T\,\psi^+(k)\,\psi(-k)>
\ee

The language proper for superconductivity will be used, and it is
assumed that the dimension is larger than 2. RPA-SDW makes the
ansatz that the spin is polarized along the Z-axis. However, I have
chosen in this section to write down Green-functions which
correspond to polarization in the XY plane, Eq. 8, to make duality
arguments more explicit.

Generically, there are three different Temperature scales present
in this problem, as shown schematically in figure 5.
The highest Temperature, denoted as $T_b$, is that at which
binding of uncorrelated pairs occurs. It is a crossover
Temperature, signaled by a
divergence in the T-matrix \cite{nozieres,jan}. A spin gap, $m_1
= 2 \Delta$,
opens up in $G_0$ because it costs some energy to break a pair.
Conversely, for $U$ positive, a
charge gap appears in $G_0$. The non-diagonal elements of
$\tilde{G}_{sup}$ and $\tilde{G}_{SDW}$,
$F(k)=<T\hat{\psi}^+(k)>$, are still zero.

At a lower temperature, $T_0$ ($\sim J$ for strong coupling),
correlations among
Cooper pairs  build up, and the correlation length $\xi(T)$ begins
to be longer than the lattice spacing, $a$. Its inverse,
$m_{corr} = 1/\xi$,
appears as a pole in the correlation function $C_{\psi}$,
which takes the conventional Ornstein-Zernike form
\be
C_{\psi}(k)=\frac{1}{\pi}\frac{m_{corr}}{k^2+m_{corr}^2}
\ee
$m_{corr}$ is the gap for excitations of the
collective modes of the order parameter. For $U$ positive,
the pole appears in the correlation function $C^{+-}$,
and $m_{corr}$ is the gap in the spectrum of spin waves
\cite{halperin}. This is also the Temperature where the anomalous
Nambu Green's functions,
$F(k\sim 1/a)=<T\hat{\psi}^+(k\sim 1/a)>$ begin to differ from zero.

Finally, at a still  lower temperature, $T_c$,
a phase transition towards
the broken symmetry superconduting state occurs, and is signaled
by a divergence of $\xi(T)$: the mass gap
$m_{corr}$ closes down, as dictated by Goldstone's theorem.
Below $T_c$, another length scale sets in: the coherence
length of Cooper pairs, which is also Josephson's
coherence length, $\xi_J=1/m_J$ \cite{josephson}:
\be
C_{\psi}(k)= 4 \pi \xi_J \frac{|\psi^*|^2}{k^2}=\frac{2 K_B
T}{\rho_s} \frac{|\psi^*|^2}{k^2}
\ee
where $\psi^*=<\hat{\psi}^+(k=0)>$ is the expectation value
of the order parameter and $\rho_s$ is the superfluid density.

Notice that for small $U$, $T_b$, $T_0$ and $T_c$ merge into one
line;
$m_1$ is then called the superconducting charge gap, and
equals $m_J$ for $T<T_c$, or $m_{corr}$ for $T>T_c$.
This means that $\xi(T)$
decreases to zero very quickly as the Temperature is raised above
$T_c$ in the weak coupling regime.
In other words, Cooper pairs become completely uncorrelated for
Temperatures slightly above $T_c$. Mean Field BCS theory is very
accurate in this limit and predicts that the order parameter
$\psi = \frac{m_1}{2 U}$.

For larger attraction $U$, these three
Temperatures depart from each other, and there is a finite
Temperature range above $T_c$ where $\xi(T)$ is different from
zero. The order parameter is no longer equal to $\frac{m_1}{2 U}$
because fluctuations become important (remember the discussion in
the former section).

Finally, for very large $U$, $T_b$
diverges, and Cooper pairs become composite bosons. $T_0$ and
$T_c$ again go to zero because it costs more
and more energy for a pair to hop from one site to the following
-the effective tunneling amplitude, $t^2/U$, vanishes-. For the
positive-U case, we would say that the spins become progressively
less correlated because the exchange integral, $J\sim 4 t^2/U$
vanishes.

\section{conclusions}
This paper shows that the RPA-SDW expansion developed in ref. 12
provides reliable results for the magnetic properties of the
Hubbard model at either weak or strong coupling even in the
extreme case of one spatial dimension. RPA-SDW does provide
a qualitative, yet not quantitative, description of the
intermediate coupling regime, when the Coulomb repulsion $U$ is
of the order of the bandwidth $D$.
These conclusions should
be quite robust against doping and Temperature
as long as the exact ground state is still Neel-like
\cite{siggia,schulz1,shastry}. The accuracy of this expansion
should improve with increasing dimensionality.
The Spin Density Wave and Quantum Neel ground states are
characterized the expectation value $<S^2>$, which measures
the degree of itinerancy of the charge degree of freedom electrons.
This criterion allows to show that the Spin Density Wave
evolves continually into the Neel state. Free electrons have
$<S^2>=3/8$, but $<S^2>$ increases quickly with $U$ and saturates to
the value of localized spins, $3/4$, when the Coulomb energy $U$
is of the order of the bandwidth.
Several Temperature and length scales are discussed in both weak
and strong coupling limits.

It is a pleasure to thank G. G\'omez-Santos, N. Lorente, J.
P\'erez-Conde, F.
Gebhard, F. Sols and E. Miranda for their helpful hints.
This work was supported by DGICYT, Project No. PB93-1248

\appendix
\section{Reminder of SDW-RPA Formalism at half-filling and
zero Temperature}

The Hubbard Hamiltonian can be rewritten in terms of the staggered
magnetization, $\hat{m}_i$
\be
\hat{\cal{H}} = \sum_{k} \epsilon_{k,\sigma}
\hat{f}_{k,\sigma}^+ \hat{f}_{k,\sigma}-
U \sum_i \hat{m}_i^2
\,=\,\hat{\cal{H}} = -2\,t \sum_{k,\sigma} \cos(k)
\hat{f}_{k,\sigma}^+ \hat{f}_{k,\sigma}-
U \sum_i \hat{m}_i^2
\ee
if we neglect charge fluctuations,
a good assumption for the half-filled case.
The staggered magnetization can, in turn, be expressed
in terms of spin-$1/2$ operators
\beq
\hat{S}_{i}^{\mu}&=&(\hat{\rho}_i,\hat{S}^x_i,\hat{S}^y_i,
\hat{S}^z_i)=
\sum_{\alpha,\beta} \hat{f}_{i,\alpha}
\tau_{\alpha\beta}^{\mu} \hat{f}_{i,\beta}\no\\
\tau^{\mu}_{\alpha\beta}&=&
\left(\delta_{\alpha\beta},
\frac{\sigma_{\alpha\beta}^{i}}{2}\right)
\eeq
as $\hat{m}_i=e^{i{\bf Q}{\bf R_i}} \hat{S}_i^z$.
A Mean Field decomposition on the interaction term
is performed now, $\hat{m}_i^2=<\hat{m}_i>\hat{m}_i$, so that
\be
\hat{\cal{H}} = \sum_{k,\sigma} \epsilon_k
\hat{f}_{k,\sigma}^+ \hat{f}_{k,\sigma}+
U <\hat{m}_i> \sum_i \hat{m}_i
\ee
where $U\,<\hat{m}_i>=\Delta$ is half the charge gap.
The Hamiltonian
is diagonalized with a
conventional Bogoliubov transformation and the
result is
\beq
\hat{\cal{H}} =
 -\sum_{k} \!'\, E_k \,v^2_k
&+& \sum_{k,\sigma} \!'\,E_k\,(\hat{c}_{k,\sigma}^+
\hat{c}_{k,\sigma}-
\hat{d}_{k,\sigma}^+ \hat{d}_{k,\sigma})\nonumber\\
E_k&=& \sqrt{\epsilon^2_k+\Delta^2}
\eeq
with the self-consistency condition
\be
\frac{1}{U}=\frac{1}{N}\sum_k\! '\,\frac{1}{E_k}
\ee
The prime in the
sum means summation in the restricted magnetic Brillouin zone. The
new ground state, $|\cal{SDW}>$, is annihilated by $\hat{c}$ and
$\hat{d}^+$.
The self-consistent  equation can be solved explicitly for large
$U$ and gives
$\Delta = U/2$, which implies that the Mean Field staggered
magnetization is $1/2$ in that limit.

We define Matsubara correlation functions as
\be
\chi^{\mu\nu}(q,q',\tau)=
\frac{i}{N}<T \,S^{\mu}(q,\tau)\, S^{\nu}(-q',0)>
\ee
It is straightforward to obtain from them the retarded
non-interacting
response functions
\beq
\chi_0^{00}(q,q',w)&
=&\delta_{q,q'}\,\sum_k \! '\left( 1-\frac{\Delta^2+\epsilon_k
\epsilon_{k+q}}{E_k
E_{k+q}}\right)
\left\{\frac{1}{w+E_{k+q}+E_{k}+i\delta}-
\frac{1}{w-E_{k+q}-E_{k}+i \delta}\right\}\nonumber\\\no\\
\chi_0^{zz}(q,q',w)&=&\frac{\chi_0^{00}(q,q',w)}{4}\no\\\no\\
\chi_0^{+\,-}(q,q',w)&=&\frac{\delta_{q,q'}}{2}\,\sum_k \!'
\left( 1+\frac{\Delta^2-\epsilon_{k} \epsilon_{k+q}}{E_{k}
E_{k+q}}\right)
\left\{\frac{1}{w+E_{k+q}+E_{k}+i\delta}-
\frac{1}{w-E_{k+q}-E_{k}+i \delta}\right\}\no\\\no\\
\chi_{Q}^{+\,-}(q,q',w)&=&\frac{\delta_{q,q'+Q}\,
\Delta}{2}\sum_k \!'\left(\frac{1}{E_{k}}+
\frac{1}{E_{k+q}}\right)
\left\{\frac{1}{w+E_{k+q}+E_{k}+i\delta}+
\frac{1}{w-E_{k+q}-E_{k}+i \delta}\right\}
\eeq
The RPA-SDW expressions are now given by
\beq
\chi_{RPA}^{00}&=&\frac{\chi_0^{00}}
{1+\frac{U}{2}\chi_0^{00}}\no\\\no\\
\chi_{RPA}^{zz}&=&\frac{\chi_0^{zz}}{1-2 \,U\,
\chi_0^{zz}}\no\\\no\\
\chi_{RPA}^{+\,-}&=&\left(\begin{array}{cc}
\chi_{o}^{+\,-}(q,w)&\chi_{Q}^{+\,-}(q,w)\\
\chi_{Q}^{+\,-}(q,w)&\chi_{0}^{+\,-}(q+Q,w)\end{array}\right)
\left(\begin{array}{cc}
1-U \chi_{o}^{+\,-}(q,w)&-U\,\chi_{Q}^{+\,-}(q,w)\\
-U\,\chi_{Q}^{+\,-}(q,w)&1-U\chi_{0}^{+\,-}(q+Q,w)
\end{array}\right)^{-1}
\eeq

The poles of
$\chi_{RPA}^{+\,-}$ give the spin-wave excitation spectrum
in the RPA-SDW approximation. They are located around the
antiferromagnetic,
$q_0=\pi$, and ferromagnetic, $q_0=0$, wave-vectors.
Both have the same velocity, but different weights.
Expanding the denominator
for small $q'=q_0 -\pi$ and $w$, the spin-wave velocity, $v_s$, in
the
$q'\rightarrow 0$ limit can be written as
\be
\frac{v_s}{t}=4\,\sqrt{\frac{y\,z\, U}{x}}
\ee
where $x$, $y$ and $z$ are given by
\beq
x&=&\sum_k\!' \frac{1}{E^3_k}\no\\
y&=&\sum_k\!'
\frac{3 \sin^2 (k)-1}{E^3_k}-12\,\sum_k\!'
\frac{\cos^2 (k) \sin^2
(k)}{E^5_k}\no\\
z&=&\sum_k\!'\frac{\cos^2(k)}{E^3_k}
\eeq
This expression for the spin-wave velocity
gives the correct asymptotics
for a half-filled band,
\beq
\frac{v_s}{t}&\rightarrow& \sqrt{d}\, J\,\,\,\,\,\,
U\rightarrow \infty\no\\
&\rightarrow&0 \,\,\,\,\,\,U\rightarrow 0\no
\eeq
where $J= \frac{4 t^2}{U}$ is the exchange integral.

The summations over the restricted Brillouin zone can be
performed analytically for large $U$. A straightforward yet tedious
calculation gives the following expression for the transverse
correlation function
\be
\chi_{RPA}^{+\,-}(q,w) = \frac{1}{U}\:\frac{w^2-w_q^2+4
(1-\cos(q))}{2\,w_q}
\left\{\frac{1}{w+w_q+i\delta}-\frac{1}{w-w_q+i\delta} \right\}
\ee
where $w_q =v_s |\sin(q)|$ is the spectrum of spin waves. The
imaginary part of the
transverse response function gives the spectral
density of these low-energy excitations:
\be
\chi_{RPA}^{+\,-}"(q,w) = \frac{1}{2} \tan(\frac{q}{2})
\delta(w-w_q)
\ee
Notice that $\chi_{RPA}^{+\,-}"(q,w)$ has a zero mode at $q=0$
with vanishing weight and another at $q=\pi$, whose weight diverges.
The real part of  the response function at zero energy
\be
\chi_{RPA}^{+\,-}"(q,0)\simeq \frac{2}{U\,v_s^2}\frac{1}{q^2}
\ee
is related to the transverse correlation function $C^{+\,-}$.

The RPA-SDW staggered magnetization can be written as
\be
<S^z>=\,<\hat{m}_i>\,=\frac{1}{2\,N}
\sum_{k,\sigma} \,\sigma\,\int_{-\infty}^{\infty}\frac{dw}{\pi}\,
A_{\sigma}(k,k+Q,w)\,n_F(w)=\frac{1}{N} \sum_k
\int_{-\infty}^{0}\frac{dw}{2\pi}\,A_{\uparrow}(k,k+Q,w)
\ee
where the spectral function $A_{\sigma}(k,k',w)$ is
obtained from Dyson's equation. The RPA-SDW self-energy is given by
the matrix equation
\be
\tilde{\Sigma}_{\sigma}(k,w)=\frac{U^2}{\beta}\sum_{k',w_B}
\tilde{G}_{\sigma}^0 (k+k',w+w_B)\:\tilde{\chi}_{-\sigma}(k',w_B)
\ee
where only the transverse and
longitudinal spin plus charge fluctuation
channels are taken into account
\cite{swz}.
It is important to realize that RPA fluctuations reduce the
staggered magnetization from its Mean Field value,
$<S^z>=\Delta/U$. In particular, for large values of $U$, Mean
Field Theory yields $<S^z>=1/2$ while RPA-SDW gives
\be
<S^z>=\frac{1}{2}\left\{1-2 \sum_{k}\,'\:\frac{1-\sin(k)}{|\sin(k)|}
\right\}\rightarrow -\infty
\ee
This result implies that one can have a charge insulating
spin disordered state, or {\em spin liquid}, within SDW-RPA
theory, i.e.: $<S^z>\leq 0$ even though $\Delta \neq 0$.

Finally, the
expectation value of $<S^2>$ is defined in terms of the
dynamic correlation function as
\be
<S^2>-<S^z>^2=\,
\sum_q\int_{-\infty}^{\infty}\frac{d\,w}{2\pi}\, C(q,w)
\ee
which, in turn, is
related with the dynamic response function through the
fluctuation-dissipation theorem which reads at $T=0$:
\be
<S^2>-<S^z>\,=\,
\sum_q \int_{0}^{\infty}\frac{d\,w}{\pi}\,
(\chi^{zz}\,"(q,w)+\chi^{+\,-}\,"(q,w))
\ee
The double prime denotes again the
imaginary part of the function and
the summation is now performed over the whole Brillouin zone.
It is interesting to
notice that the Mean Field approach
gives both the weak- and
strong-coupling limits correctly
\beq
<S^2>=S^{zz}(q,w)+S^{+\,-}(q,w)+<S^z>^2&=&
\frac{1}{8}+\frac{1}{4}=\frac{3}{8}
\,\,\,\,\,\,\,\,\,\,U= 0\no\\
&=& 0+\frac{1}{2}+\frac{1}{4}\,\,\,\,\,\,\,\,\,\,U= \infty\no
\eeq

\section{Bethe ansatz results}
The Hubbard model was solved by Lieb and Wu
(LW)\cite{lieb} in 1968.
Since then, a number of
authors have completed LW's work and obtained the different
physical magnitudes
\cite{ovchinnikov,shiba2,woynarovich,korepin,ogata}.
This appendix is devoted to display the integral
equations, the charge energy gap, $<S^2>$ and $v_s$ as
obtained  by Shiba \cite{shiba2} and
Ovchinnikov \cite{ovchinnikov}.

The integral equations at zero field
and half-filling for the distribution of the
k-rapidities are given by
\beq
\rho (k)&=&\frac{1}{2\pi}+\frac{2}{\pi} \,
\cos (k) \int_{-\pi}^{\pi} dk'\, \rho (k')\: S(2 (\sin
(k)-\sin (k'),u)\\
\partial_u \rho (k)&=&
\frac{2}{\pi u}\,\cos (k) \int_{-\pi}^{\pi}dk'\, \rho (k') \:R(2
(\sin
(k)-\sin (k'),u)+\no\\
&+&\frac{2}{\pi} \,\cos (k) \int_{-\pi}^{\pi} dk'\,
\partial_u\rho (k')\:
\{S(2(\sin (k)-\sin (k')),u)-S(2 \sin (k),u)\}\no
\eeq
where the kernels $S$ and $R$ are
\beq
S(x,u)&=& \sum_1^{\infty} (-1)^{n+1} \frac{n u}{x^2+(n u)^2}\no\\
R(x,u)&=& \sum_1^{\infty} (-1)^{n+1}
\frac{(x^2-(n u)^2) n u}{(x^2+(n u)^2)^2}
\no\eeq
The charge energy gap is given by
\be
\frac{\Delta}{t}= 16 \,u
\int_1^{\infty} dx \,\frac{\sqrt{x^2-1}}{\sinh(\frac{2\pi x}{u})}
\ee
while the velocity of spin waves
in the $q\rightarrow 0$ limit has the following
expression
\be
\frac{v_s}{t}=\frac{2\, I_1(\frac{2\pi}{u})}{I_0(\frac{2\pi}{u})}
\ee
where $I_n$ are Bessel functions of imaginary argument. Finally,
\be
<S^2>=\frac{3}{4}+3 \int_{-\pi}^{\pi}
dk\, (1+\cos (k))\:\partial_u\rho (k)
\ee

\begin{figure}
\caption{Charge gap, $2\,\Delta$ as a
function of $U$. Solid line is the
bethe ansatz exact answer;
dashed line is the RPA-SDW Mean Field result}
\end{figure}

\begin{figure}
\caption{(a)
$\chi^{+-}_{RPA}"$ as a function of $q$ and $w$ for $U = 20$; (b)
$\chi^{zz}_{RPA}"(q,w)$ for the same value of $U$}
\end{figure}

\begin{figure}
\caption{Spin-wave velocity,
$v_s$, versus $U$. Solid line is the exact
result; dashed line is RPA's answer.
Notice that $v_s$ tends to $0$ as
$U$ goes to $0$, and shows a maximum at $U\sim 1$}
\end{figure}

\begin{figure}
\caption{$<S^2>$ versus $U$. Solid line is the Bethe ansatz result
and dashed line is Mean Field's answer}
\end{figure}

\begin{figure}
\caption{Characteristic Temperature of the different phenomena as a
function of $U$. Solid line is the Binding Temperature, $T_b$; dashed
line is the Temperature, $T_{corr}$, at which correlations begin
to build up; dotted line is the transition Temperature to Long Range
Order, $T_c$}
\end{figure}


\begin{references}
\bibitem{haldane1} This argument does not hold for some systems,
whose low energy excitation spectrum is exhausted by topological
solitons.
\bibitem{caveat} Heavy Fermion superconductors might show this
kind of pairing
\end{references}
\end{document}